\documentclass[a4paper,11pt]{article}
\pdfoutput=1 

\usepackage{jheppub} 

\usepackage[T1]{fontenc} 

\usepackage{physics}

\usepackage[compat=1.1.0]{tikz-feynman}

\newcommand{\feynalign}[1]{
    \begin{gathered}
    \begin{tikzpicture}[scale=0.4, transform shape, baseline={([yshift=-.5ex]current bounding box.center)}]
    #1
    \end{tikzpicture}
    \end{gathered}
}

\newcommand{\criticalBubbleMod}{
    \coordinate (a) at (0.0,0.0);
    \coordinate (b) at (0.0,0.7);
    
    \draw (b) arc [start angle=-90, end angle=270, radius=0.7cm];
    
    \draw (a) -- (b); 
}

\newcommand{\fluctuationsMod}{
    \coordinate (a) at (-0.495,0.0);
    \coordinate (c) at (0.495,0.0);
    \coordinate (b) at (0.0,0.7);
    
    \draw (b) arc [start angle=-90, end angle=270, radius=0.7cm];
    
    \draw (a) -- (b);
    \draw (c) -- (b);
}

\newcommand{\pr}{\phi_\text{r}}
\newcommand{\pa}{\phi_\text{a}}
\newcommand{\cb}{\phi_\text{cb}}
\newcommand{\dpr}{\delta\phi_\text{r}}
\newcommand{\pe}{\phi_\text{E}}
\newcommand{\psp}{\phi_\text{sp}}
\newcommand{\dpe}{\delta\phi_\text{E}}
\newcommand{\se}{S_\text{E}}
\newcommand{\prj}{\phi_{\text{r}j}}
\newcommand{\paj}{\phi_{\text{a}j}}
\newcommand{\prjp}{\phi_{\text{r}{j_+}}}

\newcommand{\pmeta}{\phi_\text{meta}}
\newcommand{\pstable}{\phi_\text{stable}}
\newcommand{\pam}{\phi_{\text{a}-}}

\newcommand{\funcdv}[1]{\frac{\delta}{\delta#1}}
\newcommand{\funcdvsq}[1]{\frac{\delta^2}{\delta#1^2}}

\title{\boldmath Quantum field nucleating\\and Wigner functions}

\author[]{Joonas Hirvonen}

\affiliation[]{School of Physics and Astronomy,
    University of Nottingham,
    Nottingham NG7 2RD,
    U.K.}

\emailAdd{joonas.hirvonen@nottingham.ac.uk}

\abstract{We present a novel real-time framework for the decay of metastable states in quantum field theories using Wigner functions. The framework introduces a nonperturbative  nucleation rate formula that captures both quantum tunneling and over-the-barrier nucleation, alongside steps to evaluate it perturbatively. We apply it to a simple thermal example with direct relevance to current analog experiments. Our derived one-loop nucleation rate fundamentally differs from the widely cited high-temperature result by Linde: The prefactor contains quantum effects and also asymptotes to a differing form at high temperatures, where the quantum effects become negligible. Rather, the result is a generalization of Affleck's rate formula to quantum field theories, asymptoting to the effective field theory approach, and Langer's rate, at high temperatures. The example also reveals that the high-temperature side of the ``quantum-to-classical'' transition of thermal vacuum decay is still inherently quantum mechanical, even though the bounce background possesses the classical, $\mathrm{O}(d)\times S^1$, symmetry.}

\begin{document} 
\maketitle
\flushbottom

\section{Introduction}

First-order phase transitions are fascinating and ubiquitous phenomena in nature. A system is trapped in a metastable phase, and regions of a new stable phase begin to nucleate. The regions grow until the system has reached a new equilibrium, either one of phase coexistence or fully in the new phase.

Studying first-order phase transitions attracts interest particularly due to two experimental fronts: gravitational waves from early-universe phase transitions~\cite{LISA:2017pwj,NANOGrav:2020bcs} and table-top experiments~\cite{Fialko:2014xba, Fialko:2016ggg, Billam:2021nbc,Song:2021pyy, Tian:2022dzv,Viermann:2022wgw, Zenesini:2023afv, Jenkins:2023eez, Jenkins:2023npg,QUEST-DMC:2024crp, Darbha:2024srr, Zhu:2024dvz,Cominotti:2025qia,Vodeb:2024tvo,Osterholz:2025yiw,Luo:2025qlg}. It has long been known that first-order phase transitions during the cosmological evolution may have produced a gravitational wave background~\cite{Witten:1984rs,Hogan:1986dsh,Caprini:2018mtu,Caprini:2019egz,Hindmarsh:2020hop,Athron:2023xlk}. The phase transition of quantum chromodynamics~\cite{Aoki:2006we,Aoki:2006br,Aoki:2009sc,Borsanyi:2010bp,Bhattacharya:2014ara,Bazavov:2011nk} and the electroweak phase transition~\cite{Kajantie:1996mn,Kajantie:1995kf,Laine:1998vn} are both crossovers at high temperatures according to the Standard Model of particle physics. Hence, an observation of gravitational waves from an early-universe first-order phase transition would be a signal from physics beyond the Standard Model. There is already a potential candidate signal from pulsar timing arrays~\cite{Nakai:2020oit,Ratzinger:2020koh,NANOGrav:2021flc,Bringmann:2023opz,Madge:2023dxc,Goncalves:2025uwh}. The table-top experiments, on the other hand, provide a controllable environment for testing our theories of phase transition dynamics (see e.g.\ Ref~\cite{Garcia:2025uph}).

The theoretical foundations for nucleation began to take shape with Gibbs by the introduction of the critical bubble, which is a configuration on the verge of transitioning from the metastable phase to the stable phase~\cite{Gibbs1874}. There were multiple improvements to take into account the dynamics in classical nucleation theory~\cite{becker1935kinetische,wigner1938transition,zeldovich1942theory,cahn1959free3} culminating in Langer's theory of classical nucleation~\cite{Langer:1967ax,Langer:1969bc,Langer:1974cpa}. On the side of quantum theory, Coleman and Callan formulated the vacuum decay rate in quantum field theories (QFTs)~\cite{Coleman:1977py,Callan:1977pt}, Affleck found the escape rate of a quantum particle escaping a metastable well in finite temperature~\cite{Affleck:1980ac}, and Linde found a rate for a quantum field in finite temperature~\cite{Linde:1981zj}.

There has been a resurgence of research in the false vacuum decay rate in many different fronts:  direct formulations of zero-temperature vacuum decay instead of the Euclidean formulation by Coleman and Callan~\cite{Andreassen:2016cff,Ai:2019fri,Garbrecht:2025alb}, connecting nucleation in high-temperature QFTs to Langer's theory using effective field theories (EFTs)~\cite{Gould:2021ccf}, showing the gauge invariance of high-temperature nucleation rate~\cite{Hirvonen:2021zej,Lofgren:2021ogg}, extending Langer's theory to all orders in perturbation theory~\cite{Ekstedt:2022tqk}, finding real-time instantons in finite-temperature~\cite{Steingasser:2023gde,Steingasser:2024ikl}, describing real-time nucleation in high-temperature QFTs using Boltzmann equations~\cite{Hirvonen:2024rfg}, studying the validity of Langer's theory in real-time simulations~\cite{Pirvu:2024nbe,Pirvu:2024ova,Hirvonen:2025hqn}, proving the $\mathrm{O}(d)$ invariance of the thermal bounce solution~\cite{Shoji:2025nvj}, and describing decay in the presence of conserved charges~\cite{Barni:2026dhc,Barni:2026fvy}.

In this article, we present a novel first-principles framework for computing decay rates in QFTs based on Wigner functions. We show that the total decay rate is captured in a simple nonperturbative formula, which includes quantum tunneling that dominates at low temperatures and over-the-barrier nucleation that dominates at high temperatures. We also lay down the steps for evaluating the formula perturbatively for the over-the-barrier process.

We perform the computation in a real-scalar theory, applicable to analog experiments (\textit{cf.} Refs.~\cite{Zenesini:2023afv,Garcia:2025uph}), and show that the result fundamentally disagrees with Linde's nucleation rate on two accounts. Our result reveals that there can be strong quantum corrections to the prefactor that were previously omitted. In addition, the formulae do not agree even at high temperatures, where the quantum corrections become negligible. Instead, our result asymptotes to Langer's classical nucleation rate through the EFT construction of high-temperature dimensional reduction~\cite{Farakos:1994kx,Kajantie:1995dw,Braaten:1995cm,Braaten:1995jr,Hirvonen:2022jba} as expected based on Ref.~\cite{Gould:2021ccf}. The derived rate formula can be viewed as the generalization of Affleck's rate to QFTs. By a quirk in history, he never published the article on QFTs he cites in Ref.~\cite{Affleck:1980ac}, leaving Linde's formula to become ubiquitously cited across high-temperature QFT literature.

Wigner functions have been used to study false vacuum decay in Refs.~\cite{Calzetta:2001pp,Braden:2018tky,Hertzberg:2019wgx,Wang:2025ooq}. The analyses have been performed in the context of truncating the time evolution of the Wigner function in powers of $\hbar$ in regimes of non-linear evolution of the quantum field: from the metastable phase to the escape. Also, the initial state around the metastable phase is truncated. We will discuss in Appendix~\ref{app:truncatedWigner}, why we are doubtful of the results in the aforementioned references. See also Ref.~\cite{Tranberg:2022noe} for discussion on the truncated Wigner approach.

Another approach very similar to ours is to evolve an approximate classical probability distribution for longwavelength bosonic fields~\cite{Moore:2000jw,Moore:2001vf} (see also Refs.~\cite{Gould:2022ran,Batini:2023zpi,Gould:2024chm,Pirvu:2024nbe,Pirvu:2024ova,Hirvonen:2025hqn} for more recent results from and discussions on the approach). Fundamentally, this arises from the classicalization of the Wigner function for these modes at high temperatures, which we also observe in our analysis. In Ref.~\cite{Hirvonen:2024rfg}, the probability distribution was extended to contain thermal off-equilibrium particles. The extension was constructed on the basis of kinetic equations and their effective Hamiltonian (see Refs.~\cite{Blaizot:2001nr,Nair:1993rx,Iancu:1998bmf} for gauge fields). The Wigner function method allows for computing rates below the high-temperature regime. After generalizing to asymptotically high temperatures, it should also allow for going beyond the limitations of the classical probability distribution, for example to include the effects from infrared quantum fluctuations to the rate.

In Sec.~\ref{sec:definingNucleationRateWithWignerFunction}, we formulate the full decay rate based on Wigner functions. In Sec.~\ref{sec:quantumFieldNucleating}, we examine the dynamics of the over-the-barrier nucleation in real-time QFTs. We then use the Wigner function formulation to find the nucleation rate from a metastable phase that is (approximately) in equilibrium in Sec.~\ref{sec:equilibriumNucleation}. The low-temperature limit of validity of the result is discussed in Sec.~\ref{sec:lowTValidity}. In Sec.~\ref{sec:highTemperatureLimit}, we take the high-temperature limit of the rate. We show that it asymptotes to the effective field theory methods, discuss its high-temperature limit of validity, and remark on the extension to asymptotically high temperatures. Finally, we conclude in Sec.~\ref{sec:conclusions}.

\section{False vacuum decay with Wigner functions}\label{sec:definingNucleationRateWithWignerFunction}

We begin by discussing formulating the full escape rate of a system in terms of the Wigner function. The rate can be defined using the probability for the system being in the metastable phase, $P_\text{meta}$:
\begin{equation}
    \Gamma=-\frac{1}{P_\text{meta}}\dv{P_\text{meta}}{t}\,.
\end{equation}
For a quantum field, this contains both quantum tunneling out of the metastable phase and nucleation, which will be discussed in Sec.~\ref{sec:quantumFieldNucleating}.

The probability for being in the metastable phase is given by the density matrix,
\begin{equation}\label{eq:probabilityBasicDefinition}
    P_\text{meta}=\int_{\phi\in\text{meta}}\mathcal{D}\phi\,\rho[\phi,\phi]\,,
\end{equation}
where $\rho[\phi,\phi]$ is the probability density of finding the field in the configuration $\phi$. This also requires defining the metastable phase, $\phi\in\text{meta}$, in the configuration space. The region of metastable phase must engulf the fluctuations of the (approximate) equilibrium distribution around the metastable minimum and exclude the stable minimum and the fluctuations therein. Here, we do not need to be precise about the metastable phase, but below we will choose the boundary in a specific manner for perturbative nucleation rate computation. (See Refs.~\cite{Hirvonen:2025hqn,Moore:2000jw,Dutka:2025ghb} for discussions for defining the metastable phase.)

The probability for being in the metastable phase can also be formulated in terms of the Wigner function~\cite{Wigner:1932eb}:
\begin{align}
    P_\text{meta}&=\int_{\pr\in\text{meta}}\mathcal{D}\pr\mathcal{D}\pi W[\pr,\pi]\,,\label{eq:WignerFunctionProbabilityOfMeta}\\
    W[\pr,\pi]&=\frac{1}{\mathcal{N}}\int\mathcal{D}\pa\, \rho\qty[\pr+\frac{\pa}{2},\pr-\frac{\pa}{2}] e^{-i\int\dd^d \mathbf{x}\pi\pa}\,.\label{eq:WignerFunction}
\end{align}
The normalization factor $\mathcal{N}$ normalizes the total probability to one. The initial definition for $P_\text{meta}$ in Eq.~\eqref{eq:probabilityBasicDefinition} can be recovered by inserting the definition of the Wigner function, Eq.~\eqref{eq:WignerFunction}, in Eq.~\eqref{eq:WignerFunctionProbabilityOfMeta} and integrating over the variables $\pi$ and $\pa$. Here, we have denoted the average field by $\pr$ and the difference field by $\pa$ in anticipation of the real-time formalism used below in Sec.~\ref{sec:quantumFieldNucleating}.

The Wigner function obeys the Moyal equation~\cite{Moyal:1949sk}, which simplifies to a local continuity equation in the configuration space after integrating over the conjugate momentum,
\begin{equation}\label{eq:nonperturbativeflow}
    \int\mathcal{D}\pi\partial_t W[\pr,\pi] = - \int\mathcal{D}\pi\,\pi\cdot\funcdv{\pr} W[\pr,\pi]\,.
\end{equation}
(See Appendix~\ref{app:WignerFunctionTimeEvolution}.) Here, the dot product is defined as
\begin{equation}\label{eq:dotProduct}
    \pi\cdot\funcdv{\pr}\equiv\int\dd^d\mathbf{x}\,\pi(\mathbf{x})\funcdv{\pr(\mathbf{x})}\,.
\end{equation}
Due to the continuity equation, the escape rate becomes simply an integral over the boundary of the metastable phase,
\begin{equation}
    \Gamma =\frac{\int\mathcal{D}\pi\int_{\partial(\pr\in\text{meta})}\mathcal{D}S\cdot\pi W[\pr,\pi]}{\int_{\pr\in\text{meta}}\mathcal{D}\pr\mathcal{D}\pi W[\pr,\pi]}\,,\label{eq:nonperturbativeRate}
\end{equation}
where $\mathcal{D}S$ is the area element normal to the surface, pointing outwards from the metastable phase.

The above result may look deceptively classical. However, it is completely general for quantum field theory; no approximation has been made. The quantum nature lies in the Wigner function $W[\pr,\pi]$, which is not a classical probability distribution, often referred to as the Wigner quasiprobability distribution, which can also have negative values. Integrating over the $\pi$ variable in the Wigner function does however result in the probability density for $\pr$: $\rho[\pr,\pr]$.

\section{Quantum field nucleating over the barrier}\label{sec:quantumFieldNucleating}

Here, we begin to focus on the case of a quantum field escaping the metastable phase by evolving over the barrier. We overview how to obtain a real-time semiclassical equation of motion for the quantum field around the critical bubble, which is a saddle point configuration between the metastable phase and the stable phase. This is performed in the closed time path (CTP) formalism (also known as the Schwinger-Keldysh or in-in formalism)~\cite{Schwinger:1960qe,Keldysh:1964ud}. We will also find a suitable surface for a perturbative evaluation of the rate formula in Eq.~\eqref{eq:nonperturbativeRate} for the over-the-barrier process. Physically, we also want to highlight the existence of the over-the-barrier process as a separate decay channel to quantum tunneling, even when the quantum field is not behaving approximately classically due to thermal effects in high temperatures.

The semiclassical equation of motion around the critical bubble arises from treating nonlinearities perturbatively. We will find the limits of validity for the procedure in Secs.~\ref{sec:lowTValidity} and \ref{sec:highTemperatureLimit}, where the metastable phase has been assumed to be approximately in equilibrium (see Sec.~\ref{sec:equilibriumNucleation}). The low-temperature limit of validity results from strong quantum fluctuations in Sec.~\ref{sec:lowTValidity} and the high-temperature limit from strong thermal fluctuations in Sec.~\ref{sec:highTemperatureLimit}. The region of validity of the procedure roughly corresponds to intermediate temperatures around the mass of the nucleating field, $T\sim m$, but is described more accurately in the aforementioned sections.

The CTP path integral is given by
\begin{equation}
    \int\mathcal{D}\phi_1\mathcal{D}\phi_2 \rho[\phi_1(t_0),\phi_2(t_0)] e^{i(S[\phi_1]-S[\phi_2])}\,,
\end{equation}
where $\rho$ is the density matrix at some initial time $t_0$. The path integral is commonly used to compute unequal-time (and equal-time) correlation functions and to handle quantum field theories out of equilibrium in real time. (See e.g.\ Refs.~\cite{Bellac:2011kqa,Laine:2016hma} for textbooks and Refs.~\cite{Ghiglieri:2020dpq,Blaizot:2001nr} for review articles.) Here, we will use it for finding nucleation from a quantum field theory.

The action for a single real scalar field is given by
\begin{equation}\label{eq:aSimpleAction}
    S[\phi] = \int\dd^D x\qty(\frac{1}{2}\partial_\mu\phi\partial^\mu\phi-V(\phi))\,.
\end{equation}
We will assume that the potential $V$ has a local minimum $V'(\pmeta)=0$ corresponding to a metastable phase and a global minimum $V'(\pstable)=0$ corresponding to a stable phase. Note that the time integration in the action is restricted to $t>t_0$. There are of course counterterms that cancel vacuum divergences, but we will not handle them explicitly. The role of the one-loop counterterms is solely to cancel against the divergence of the one-loop prefactor in the nucleation rate result in Eq.~\eqref{eq:theResult}~\cite{Callan:1977pt}.

Let us now proceed with finding the real-time equation of motion by rotating to the ``retarded--advanced'' basis,
\begin{equation}\label{eq:retardedAdvanceBasis}
    \pr = \frac{\phi_1+\phi_2}{2}\,,\qquad\qquad
    \pa = \phi_1-\phi_2\,,
\end{equation}
and integrating by parts.
The exponent takes the following form
\begin{align}\label{eq:ApproximatingAwayThePA}
    S[\phi_1]-S[\phi_2]&=\int\dd^D x\Big(-\pa\big(\Box\pr+V'(\pr)\big)+\order{\pa^3}\Big)-\int\dd^d \mathbf{x}\pa(t_0)\partial_t\pr(t_0)\,,
\end{align}
where the nonlinear terms in $\pa$ are in $\order{\pa^3}$ and the second term has appeared as a boundary term from integrating by parts.

There is a time-independent stationary point in the path integral that corresponds to the critical bubble
\begin{equation}\label{eq:criticalBubble}
    -\grad^2\cb+V'(\cb)=0\,,\qquad \pr=\cb\,.
\end{equation}
Let us expand around the critical bubble, $\pr=\cb+\dpr$, for $t>t_0$:
\begin{align}\label{eq:expandedAction}
    S[\phi_1]-S[\phi_2]&=\int\dd^D x\Big(-\pa\big(\Box+V''(\cb)\big)\dpr+\order{\dpr^2\pa,\pa^3}\Big)-\int\dd^d \mathbf{x}\pa(t_0)\partial_t\pr(t_0)\,.
\end{align}

We can integrate over the advanced field at times greater than the initial time, $\pa(t>t_0)$:
\begin{align}
    &\int\mathcal{D}\phi_1\mathcal{D}\phi_2 \rho[\phi_1(t_0),\phi_2(t_0)] e^{i(S[\phi_1]-S[\phi_2])}\nonumber\\
    \approx&\int\mathcal{D}\pa\mathcal{D}\pr\mathcal{D}\dpr \rho\qty[\pr+\frac{\pa}{2},\pr-\frac{\pa}{2}] e^{-i\int\dd^d \mathbf{x}\pa\partial_t\pr} \prod_{\mathbf{x},t>t_0}\delta\qty(\frac{(\Box+V''(\cb))\dpr}{2\pi})\,.
\end{align}
We have suppressed the $t_0$ argument on the second line from $\pr$ and $\pa$, which will now refer to the initial-time fields, $\pr(t_0),\pa(t_0)$, to avoid notational clutter. The variable $\delta\pr$ still depends on time.

The path integral now gives a linear equation of motion that describes real-time over-the-barrier nucleation of the quantum field:
\begin{equation}\label{eq:EoM}
    \big(\Box+V''(\cb)\big)\dpr=0\,.
\end{equation}

Finally, we note that the initial conditions to the equation motion at $t=t_0$ are given by the Wigner function: Replacing the density matrix with the inverted Wigner transformation of the Wigner function (\textit{cf}.\ Eq.~\eqref{eq:WignerFunction})
the path integral becomes
\begin{align}
    &\int\mathcal{D}\phi_1\mathcal{D}\phi_2 \rho[\phi_1(t_0),\phi_2(t_0)] e^{i(S[\phi_1]-S[\phi_2])}\nonumber\\
    \approx&\int\mathcal{D}\pr\mathcal{D}\pi\mathcal{D}\dpr W[\pr,\pi] \prod_\mathbf{x}\delta\qty(\frac{\pi-\partial_t\pr}{2\pi}) \prod_{\mathbf{x},t>t_0}\delta\qty(\frac{(\Box+V''(\cb))\dpr}{2\pi})\,.\label{eq:finalFormRealTimeContour}
\end{align}
The first delta function enforces the initial conditions of $\partial_t\pr=\pi$.

Now, we can also find a meaningful boundary to the metastable phase around the barrier. It will also be convenient for the perturbative evaluation of the nucleation rate formula in Eq.~\eqref{eq:nonperturbativeRate} regarding the over-the-barrier escape. We choose the surface to go through the critical bubble, $\cb$ in Eq.~\eqref{eq:criticalBubble}. There is one unstable direction in the semi-classical equation of motion in Eq.~\eqref{eq:EoM}, which corresponds to the negative eigenmode of the critical bubble, $f_-$:
\begin{gather}\label{eq:negativeEigenMode}
    (-\grad^2\cb+V''(\cb))f_-=\lambda_-f_-\,.
\end{gather}
The boundary in the vicinity of the critical bubble can be defined to be perpendicular to the unstable direction:
\begin{gather}
    \phi_-=0\,,\\
    \phi_-:=\int\dd^d \mathbf{x}f_-(\pr-\cb)\,.\label{eq:defNegEigCoeff}
\end{gather}

The nonperturbative nucleation rate formula in Eq.~\eqref{eq:nonperturbativeRate} becomes
\begin{equation}\label{eq:agnosticNucleationRateFormula}
    \Gamma=\frac{\int\mathcal{D}\pr\mathcal{D}\pi W[\pr,\pi]\, \pi_-\delta(\phi_-)}{\int_{\pr\in\text{meta}}\mathcal{D}\pr\mathcal{D}\pi W[\pr,\pi]}
\end{equation}
with this choice of surface.
The negative eigenmode coefficient for the conjugate momentum is given by
\begin{equation}
    \pi_-=\int\dd^d \mathbf{x}f_-\pi\,.
\end{equation}

We would like to clarify now two key distinctions between our analysis and the conventional truncated Wigner approximation: We have not performed any approximation regarding the initial state Wigner function, $W[\pr,\pi]$, in Eq.~\eqref{eq:finalFormRealTimeContour}. However, the time-evolution that we obtained is limited in validity to describing the dynamics around the critical bubble, $\cb$. The former distinction is important because, as we will observe, the nucleation occurs from the highly non-Gaussian tail of the quantum distribution around the metastable state, which we need to retain. The latter one states that we cannot describe the evolution of the Wigner function everywhere in the phase space, only close to the critical bubble. Consequently, we must infer the rate from this part of the phase space. This is indeed possible possible with the rate formula in Eq.~\eqref{eq:agnosticNucleationRateFormula}, accompanied by the equilibrium assumption presented in Sec.~\ref{sec:equilibriumNucleation}.

Note the despite us evaluating the Wigner function using the ansatz of the metastable phase being in equilibrium, the formula in Eq.~\eqref{eq:agnosticNucleationRateFormula} is actually agnostic to the state and can be used for over-the-barrier nucleation with any state, even a time-dependent one, as long as the Wigner function is known on the surface.

\section{Quantum field nucleating from equilibrium}\label{sec:equilibriumNucleation}

Here, we will study the over-the-barrier escape further by assuming that nucleation occurs from the metastable phase that is in equilibrium. This will give us an explicit one-loop nucleation rate result.

The conditions for nucleation from the metastable in equilibrium phase can be formulated in terms of the Wigner function:
\begin{align}
    W_\text{nucl}[\pr,\pi] &= \sigma[\pr,\pi]\times W_\text{eq}[\pr,\pi]\,,\label{eq:nucletaionFormWignerFunction}\\[3pt]
    \sigma[\pr,\pi] &\approx 1\,,\qquad\qquad \phi\approx \pmeta\,,\\
    \sigma[\pr,\pi] &\approx 0\,,\qquad\qquad \phi\approx \pstable\,.
\end{align}
(See Refs.~\cite{Kramers:1940zz,Langer:1969bc} for classical treatments.)
The Wigner function $W_\text{eq}$ corresponds to total equilibrium (but normalized to the metastable state), and the non-trivial off-equilibrium factor $\sigma$ satisfies the Moyal equation and sets the appropriate boundary conditions.
Note that the Wigner function is time-independent, i.e.\ stationary, due to the nucleation being sourced by the equilibrium in the metastable phase.

The strategy we have constructed for obtaining the nucleation rate can be summarized as follows.
\begin{enumerate}
    \item Find the equilibrium density matrix around the critical configuration to a desired order from the imaginary time formalism. \label{i:step1}
    \item Wigner transform the equilibrium density matrix to obtain the equilibrium Wigner function, Eq.~\eqref{eq:WignerFunction}.
    \item Separate the part that solves the Moyal equation (to the desired order) and has the following boundary conditions: equilibrium in the metastable phase, zero in the stable phase.
    \item Compute the nucleation rate with Eq.~\eqref{eq:agnosticNucleationRateFormula}. \label{i:step4}
\end{enumerate}
We will perform this to the one-loop order below.

We begin from the Euclidean path-integral for the density matrix
(see e.g.\ Ref.~\cite{Laine:2016hma}):
\begin{align}
    \rho_\text{eq}[\phi,\phi']&=Z^{-1}\bra{\phi}e^{-\beta\hat{H}}\ket{\phi'}=Z^{-1}\int_{\pe(\tau=0)=\phi}^{\pe(\tau=\beta)=\phi'}\mathcal{D}\pe\, e^{-\se}\,, \label{eq:equilibriumDensityMatrix}\\
    \se&=\int_0^\beta\dd\tau\int\dd^d\mathbf{x}\qty(\frac{1}{2}(\partial_\tau\pe)^2-\frac{1}{2}\pe\grad^2\pe+V(\pe))\,.
\end{align}

We are interested in the density matrix around $\pr=\cb$, $\pa=0$, since this corresponds to the dynamics studied in  Sec.~\ref{sec:quantumFieldNucleating} (see Eqs.~\eqref{eq:criticalBubble}, \eqref{eq:expandedAction}, \eqref{eq:agnosticNucleationRateFormula}). This further corresponds to the saddle point of
\begin{equation}\label{eq:EuclideanSaddlePoint}
    \pe(\tau,\mathbf{x}) = \cb(\mathbf{x})\,.
\end{equation}

Note that other possible saddle points contribute to the density matrix additively:
\begin{equation}
    \rho_\text{eq}\approx \rho_\text{cb}+\sum_{\substack{\text{other saddle}\\ \text{points}}} \rho_{\text{sp},i}\,.
\end{equation}
Due to the linearity of the Wigner transformation, Eq.~\eqref{eq:WignerFunction}, and the nonperturbative nucleation rate formula, Eq.~\eqref{eq:nonperturbativeRate}, we will obtain at least an additive component of the full nucleation rate: the one that survives to high temperatures (see Sec.~\ref{sec:highTemperatureLimit}).

It is in some sense obvious that the background $\pe(\tau,\mathbf{x}) = \cb(\mathbf{x})$ corresponds to the physics in Sec.~\ref{sec:quantumFieldNucleating}, as noted above. Let us still elaborate this a bit further: The values at $\tau=0$ and $\tau=\beta$ yield neatly
\begin{align}
    \pr&=\frac{\phi+\phi'}{2}=\frac{\pe(\tau=0)+\pe(\tau=\beta)}{2}=\cb\,,\\
    \pa&=\phi-\phi'=\pe(\tau=0)-\pe(\tau=\beta)=0\,,
\end{align}
which do correspond to the over-the-barrier nucleation described in Sec.~\ref{sec:quantumFieldNucleating}. The saddlepoint is governed by a second-order differential equation. Hence, the values at the boundaries may not fully constrain the saddlepoint. If we allow for non-zero Euclidean time derivatives at the endpoints for the saddlepoint solution, $\psp$, the action contains linear terms for $\pr$ and $\pa$:
\begin{align}
    \se[\psp+\dpe]&\supset\int\dd^d\mathbf{x}\,\dpe\partial_\tau\psp\eval_{\tau=0}^{\tau=\beta}\\
    &=\int\dd^d\mathbf{x}\qty[\dpr\Big(\partial_\tau\psp(\beta)-\partial_\tau\psp(0)\Big)-\pa\frac{\partial_\tau\psp(\beta)+\partial_\tau\psp(0)}{2}]\,.
\end{align}
These linear terms will push the Gaussian integral away from $\pe=\cb$ or $\pa=0$. Thus, we require $\partial_\tau\psp(0)=\partial_\tau\psp(\beta)=0$, which leads to the saddle point in Eq.~\eqref{eq:EuclideanSaddlePoint}.%
\footnote{
This argument does not actually constrain the derivatives completely. The derivatives can be proportional to the negative eigenmode in Eq.~\eqref{eq:negativeEigenMode}: $\partial_\tau\pe(0,\mathbf{x})=-\partial_\tau\pe(\beta,\mathbf{x})= A\, f_-(\mathbf{x})$. This only produces a linear term for $\pr$ perpendicular to the integration surface in Eq.~\eqref{eq:agnosticNucleationRateFormula}. Hence, the flow over the barrier at the surface would still be describable by the discussion in Sec.~\ref{sec:quantumFieldNucleating}. However, a single degree of freedom, $A$, generically does not allow the infinitely many boundary conditions to be satisfied for the saddle point.
}

The equilibrium Wigner function around the critical bubble is compute explicitly in Appendix~\ref{app:oneloopEquilibriumWignerFunction}. The result is
\begin{equation}\label{eq:equilibriumWignerFunctionMainText}
    W_\text{cb}[\cb+\pr,\pi]=\frac{e^{-\beta E_\text{cb}}}{\tilde{Z}}\prod_j\frac{1}{\sqrt{2}\cosh(\beta \sqrt{\lambda_j}/2)} e^{-\tanh(\frac{\beta\sqrt{\lambda_j}}{2})\qty(\sqrt{\lambda_j}\prj^2+\frac{\pi_j^2}{\sqrt{\lambda_j}})}
\end{equation}
(see Eq.~\eqref{eq:equilibriumWignerFunctionAppendix} below). The exponential suppression comes from the Euclidean action evaluated on the critical bubble:
\begin{equation}
    \beta E_\text{cb} = \int_0^\beta\dd\tau\int\dd^d\mathbf{x}\qty(-\frac{1}{2}\cb\grad^2\cb+V(\cb))\,.
\end{equation}
The eigenmodes, $\lambda_j$, and coefficients, $\prj,\pi_j$, correspond to the orthonormal eigenmodes given by
\begin{gather}
    \qty(-\grad^2+V''(\cb))f_j=\lambda_jf_j\,,\label{eq:fluctOperatorOnGenEigMode}\\
    \pr=\cb + \sum_j \prj f_j\,,\\
    \pi = \sum_j\pi_j f_j\,.
\end{gather}
Note that the result here is analytically continued into a neat form regarding the negative and zero eigenmodes. The non-continued result is explicitly shown below in Eq.~\eqref{eq:WignerFunctionNoAnalyticContinuation}.

We can normalize the Wigner function corresponding to the metastable phase:
\begin{equation}
    \tilde{Z}_\text{meta}=e^{-\beta E_\text{meta}}\prod_j\frac{\pi}{\sqrt{2}\sinh(\frac{\beta\sqrt{\lambda_j^\text{meta}}}{2})}\,.
\end{equation}
The eigenvalues and the energy are defined similarly to above but around the metastable phase, $\pr=\pmeta$, instead of the critical bubble, $\pr=\cb$.
(See Eq.~\eqref{eq:metastableNormalizationPartitionFunction} in Appendix~\ref{app:oneloopEquilibriumWignerFunction}.)
The nucleation rate is normalized with the probability of being in the metastable phase, Eq.~\eqref{eq:nonperturbativeRate}. Hence, the overall normalization of the Wigner function cancels. Due to normalizing the Wigner function to the metastable phase, the denominator in the nucleation rate formula is simply one.

The off-equilibrium factor, $\sigma$, in Eq.~\eqref{eq:nucletaionFormWignerFunction} takes the simple form of
\begin{equation}\label{eq:stepForNucleation}
    \sigma[\pr,\pi]=\theta(\pi_--\sqrt{\abs{\lambda_-}}\phi_-)\,.
\end{equation}
It both ensures the appropriate boundary conditions for nucleation and satisfies the leading-order Moyal equation,
\begin{equation}\label{eq:LOMoyal}
    \partial_t W[\pr,\pi]=\int\dd^d\mathbf{x}\qty(-\pi\funcdv{\pr}+\Big(\qty(-\grad^2+V''(\cb))\pr\Big)\funcdv{\pi})W[\pr,\pi]\,,
\end{equation}
derived in Appendix~\ref{app:WignerFunctionTimeEvolution}, Eq.~\eqref{eq:LiouvilleEquation}.

We can already fully understand the form of $\sigma$ based on the semi-classical equation of motion in Eq.~\eqref{eq:EoM}: The argument of the step function retains its sign under the time evolution due to only decreasing exponentially in time,
\begin{equation}
    \dv{t}\qty(\pi_--\sqrt{\abs{\lambda_-}}\phi_-)=-\sqrt{\abs{\lambda_-}}\qty(\pi_--\sqrt{\abs{\lambda_-}}\phi_-)\,.
\end{equation}
Hence, the step function separates the phase-space trajectories into the ones originating from the metastable phase ($\theta=1$) and the ones originating from the stable phase ($\theta=0$), and only picks up the trajectories from the metastable phase due to the positive argument.

The Wigner function corresponding to nucleation becomes
\begin{equation}\label{eq:CarvedOutWignerFunction}
    W_\text{nucl}[\pr,\pi]=e^{-\beta \Delta E_\text{cb}}\theta(\pi_--\sqrt{\abs{\lambda_-}}\phi_-)\prod_j\frac{\sinh(\frac{\beta\sqrt{\lambda_j^\text{meta}}}{2})}{\pi\cosh(\frac{\beta\sqrt{\lambda_j}}{2})} e^{-\tanh(\frac{\beta\sqrt{\lambda_j}}{2})\qty(\sqrt{\lambda_j}\prj^2+\frac{\pi_j^2}{\sqrt{\lambda_j}})}\,,
\end{equation}
where the energy is defined as $\Delta E_\text{cb}=E_\text{cb}- E_\text{meta}$.

The off-equilibrium factor, $\sigma$, in Eq.~\eqref{eq:stepForNucleation} takes a simple and physically transparent form. By Wigner transforming the full nucleating Wigner function, $W_\text{nucl}$, one can find the analogous off-equilibrium factor for the density matrix, $\rho_\text{nucl}$:
\begin{equation}
    \sigma[\phi,\phi']=\frac{1-\erf\qty[\sqrt{\abs{\lambda_-}\tan(\frac{\beta\abs{\lambda_-}}{2})}\,\qty(\phi_--\frac{i}{2}\cot(\frac{\beta\abs{\lambda_-}}{2})\pam)]}{2}\,.
\end{equation}
The relative simplicity of the results is striking and highlights the elegance of using Wigner functions to find the state, in addition to the simple nucleation rate formula in Eq.~\eqref{eq:nonperturbativeRate}.

Before computing the nucleation rate, let us now showcase the non-continued version:
\begin{align}\label{eq:WignerFunctionNoAnalyticContinuation}
    W_\text{nucl}[\pr,\pi]&=e^{-\beta \Delta E_\text{cb}}\prod_{j}\sinh(\frac{\beta\sqrt{\lambda_{j}^\text{meta}}}{2})\\
    &\quad\times\frac{\theta(\pi_--\sqrt{\abs{\lambda_-}}\phi_-)}{\pi\cos(\frac{\beta\sqrt{\abs{\lambda_-}}}{2})} \,e^{-\tan(\frac{\beta\sqrt{\abs{\lambda_-}}}{2})\qty(-\sqrt{\abs{\lambda_-}}\phi_-^2+\frac{\pi_-^2}{\sqrt{\abs{\lambda_-}}})}\label{eq:negEigModePartOfNuclWigner}\\
    &\quad\times\prod_{j_0}\frac{1}{\pi}\,e^{-\frac{\beta\pi_{j_0}^2}{2}} \\
    &\quad\times\prod_{j_+}\frac{1}{\pi\cosh(\frac{\beta\sqrt{\lambda_{j_+}}}{2})} \,e^{-\tanh(\frac{\beta\sqrt{\lambda_{j_+}}}{2})\qty(\sqrt{\lambda_{j_+}}\prjp^2+\frac{\pi_{j_+}^2}{\sqrt{\lambda_{j_+}}})}\,.
\end{align}
Here, $j_0$ corresponds to the $d$ translational zero modes and $j_+$ to the positive eigenmodes.

Now, we can evaluate the nucleation rate in Eq.~\eqref{eq:agnosticNucleationRateFormula} to obtain our result:
\begin{align}
    \Gamma&=\frac{\sqrt{\abs{\lambda_-}}}{2\pi\sin(\frac{\sqrt{\abs{\lambda_-}}}{2T})} V\qty(\frac{2T\Delta E_\text{cb}}{\pi})^\frac{d}{2} \frac{\prod_{j}\sinh(\frac{\sqrt{\lambda_{j}^\text{meta}}}{2T})}{\prod_{j_+}\sinh(\frac{\sqrt{\lambda_{j_+}}}{2T})} e^{-\beta \Delta E_\text{cb}}\label{eq:theResult}\\
    &=\frac{V}{2\pi}\qty(\frac{\Delta E_\text{cb}}{2\pi T})^\frac{d}{2} \qty(\frac{\det(-\partial_\mu^2+V''(\pmeta))}{\det^+(-\partial_\mu^2+V''(\cb))})^{\frac{1}{2}} e^{-\beta \Delta E_\text{cb}}\,.\label{eq:theResultDeterminant}
\end{align}
The translational zero modes have been handled via the collective coordinates~\cite{GERVALS1976281,Vainshtein:1981wh} to the one-loop order. (See Refs.~\cite{Andreassen:2016cvx,Andreassen:2017rzq,Ekstedt:2022tqk} for recent discussions and higher-order corrections in vacuum decay and classical nucleation.) The differential operators on the second line are defined on the Euclidean spacetime and have periodic boundary conditions in the Euclidean time. This corresponds to having the even $n$ eigenvalues of Eq,~\eqref{eq:EuclideanEigenvalues} below. The superscript $+$ refers to omitting the one negative eigenvalue and $d$ zero eigenvalues, i.e.\ only retaining the positive ones.

The (hyperbolic) trigonometric functions contain the quantum nature of the nucleation. They result from the nonlinear tail of the quantum mechanical equilibrium distribution around the metastable phase. These are not present in Linde's rate formula~\cite{Linde:1981zj}, but are in Affleck's result~\cite{Affleck:1980ac}. The high-temperature behavior and the intricacies of the classicalization of the result are discussed in Sec.~\ref{sec:highTemperatureLimit} below, and the dominance of quantum effects at low temperatures in Sec.~\ref{sec:lowTValidity}.

Finally, we note that the assumption of a fully thermal metastable phase may not always be a totally innocent one. The metastable phase may not equilibrate quickly enough to replenish the configurations that are ripe for nucleation during the phase transition. This has been studied in classical field theories in 1+1 dimensions~\cite{Pirvu:2024nbe,Pirvu:2024ova,Hirvonen:2025hqn}. A discrepancy of approximately an order of magnitude was observed in the absence of thermal noise in some perturbative benchmark points~\cite{Pirvu:2024nbe,Pirvu:2024ova}. It is currently not clear how the results generalize to higher dimensions, where thermalization is faster, or to very different parts of the parameter/theory space. Note that the perturbative result coincides with the numerical simulations if the metastable phase is thermal~\cite{Hirvonen:2025hqn}. Additionally, the quantum nature of the real-time evolution from the metastable phase to nucleation could affect the validity of the assumption at temperatures where the nucleating degrees of freedom are not classical. Reference~\cite{Lin:2025wgc} analyzed a quantum particle in one dimension at zero temperature, but a thermal field theory analysis has not been performed due to numerical complexity.

\section{Low-temperature limit of validity}\label{sec:lowTValidity}

Let us now look at the low-temperature limit of validity. We will identify two distinct temperatures having divergences. Both are related to the fluctuations on the Euclidean interval, described by the quadratic action:
\begin{align}
    \se[\cb+\dpe]\approx\beta E_\text{cb} +\int_0^\beta\dd\tau\int\dd^d\mathbf{x}\frac{1}{2}\qty((\partial_\tau\dpe)^2+\dpe(-\grad^2+V''(\cb))\dpe)\,.
\end{align}
It is instructive to know the eigenspectrum of the fluctuations operator,
\begin{align}
    F_{jn}(\tau,\mathbf{x})&=f_j(\mathbf{x})e^{in\pi T \tau}\,,\qquad n\in\mathbb{Z}\,,\\
    \Big(-\partial_\tau^2-\grad^2+V''(\cb)\Big)\,F_{jn}&=\Big((n\pi T)^2+\lambda_j\Big)\,F_{jn}\,,\label{eq:EuclideanEigenvalues}
\end{align}
because the divergences coincide with a positive eigenmode becoming a zero mode. The eigenfunctions $f_j$ are defined in Eq.~\eqref{eq:fluctOperatorOnGenEigMode}. Note that the anti-periodic eigenfunctions of odd $n$ are included because the Wigner function is based of the full density matrix, not just its trace.

The explicit rate formula in Eq.~\eqref{eq:theResult} diverges at
\begin{equation}
    T = \frac{\sqrt{\abs{\lambda_-}}}{2\pi}
\end{equation}
due to the $\sin^{-1}(\frac{\sqrt{\abs{\lambda_-}}}{2T})$ factor. In terms of the Euclidean fluctuations, this corresponds to the lowest periodic eigenmodes, $F_{-\,\pm2}=f_- e^{\pm i2\pi T \tau}$, becoming zero modes.

Although the result suggests that the above temperature is the lower limit of validity, it is far from obvious based on the Wigner function analysis in Sec.~\ref{sec:equilibriumNucleation}. The negative eigenmode factor of the Wigner function in Eq.~\eqref{eq:negEigModePartOfNuclWigner} already diverges at twice the above temperature,
\begin{equation}
    T = \frac{\sqrt{\abs{\lambda_-}}}{\pi}\,,
\end{equation}
in two different places: $\tan(\frac{\beta\sqrt{\abs{\lambda_-}}}{2})$ in the exponent and $\cos^{-1}(\frac{\beta\sqrt{\abs{\lambda_-}}}{2})$ in the prefactor.

There are actually two problems at this temperature:
\begin{enumerate}
    \item The loop corrections to the equilibrium density matrix (and correspondingly to the Wigner function) become uncontrollable.
    \item The quantum fluctuations of $\pa$ become unsuppressed, and so the equation of motion in Eq.~\eqref{eq:EoM} is invalid.
\end{enumerate}
Both of these originate from the anti-periodic eigenmodes $F_{-\,\pm1}$ becoming zero modes. The first problem is self evident; there is now a zero mode on the Euclidean interval, which breaks the perturbation expansion.%
\footnote{
When integrating over the Euclidean fluctuations, the boundary conditions are given by $\phi$ and $\phi'$ in Eq.~\eqref{eq:equilibriumDensityMatrix}. Hence, there is only one zero mode for a real scalar field: $F_{-\,+1}-F_{-\,-1}$. See Eq.~\eqref{eq:eigenSpectrumOnTheInterval} in Appendix~\ref{app:oneloopEquilibriumWignerFunction}.
}
Hence, we cannot trust the step~\ref{i:step1} of our algorithm for finding the nucleation rate near $T = \frac{\sqrt{\abs{\lambda_-}}}{\pi}$. In terms of the field variables of $\pr$ and $\pa$, the new zero modes mean that the $\pa$ fluctuations become unsuppressed, leading to strong quantum fluctuations. (See the density matrix in Eq.~\eqref{eq:equilibriumDensityMatrixOnCBinApp} in Appendix~\ref{app:oneloopEquilibriumWignerFunction}.) Near this divergence, we can no longer neglect the $\pa$ non-linearities (\textit{cf.} e.g.\ Eq.~\eqref{eq:ApproximatingAwayThePA}). The quantum effects becoming strong means that we cannot approximate the Moyal equation with the classical Liouville equation to leading order. Worse still, at $T < \frac{\sqrt{\abs{\lambda_-}}}{\pi}$, the critical bubble is no longer a minimum of the Euclidean action, $\pe(\tau,\mathbf{x})=\cb(\mathbf{x})$, due to $F_{-\,\pm1}$ becoming negative modes.

With such a complete breakdown of the method, how could one trust the result for $\frac{\sqrt{\abs{\lambda_-}}}{2\pi}<T<\frac{\sqrt{\abs{\lambda_-}}}{\pi}$? \textit{If} the nucleation rate diverges with the periodic eigenmodes, then we can argue this based on the uniqueness of asymptotic expansions: We have found the first term for the asymptotic expansion for the over-the-barrier nucleation for $T>\frac{\sqrt{\abs{\lambda_-}}}{\pi}$. Due to the uniqueness, its continuation also holds for lower temperatures if there is no divergence at $T=\frac{\sqrt{\abs{\lambda_-}}}{\pi}$.

Evidence for the rate changing only at the lower temperature comes from Ref.~\cite{Affleck:1980ac}, in which Affleck found the full temperature dependence for the escape rate of a particle in one spatial dimension. The Wigner function analysis would have the same divergence there, but the result approximately holds down to $T=\frac{\sqrt{\abs{\lambda_-}}}{2\pi}$. A Wigner function analysis to two loops would clarify the lower limit of validity for QFTs: Does the two-loop result diverge at  $T=\frac{\sqrt{\abs{\lambda_-}}}{\pi}$, or only at $T=\frac{\sqrt{\abs{\lambda_-}}}{2\pi}$?

\section{High-temperature limit}\label{sec:highTemperatureLimit}

In this section, we will discuss the high-temperature limit of the nucleation rate obtained in Sec.~\ref{sec:equilibriumNucleation}. We will see how it relates to Langer's rate via the construction of an EFT. Therefore, the result also fundamentally disagrees with Linde's formula for high-temperature rate even when the quantum nature of the fluctuations becomes negligible. The EFT construction becomes mandatory at some high temperature, $T\gg m$, giving an upper limit of validity for the rate formula in Eq.~\eqref{eq:theResult}. We will also briefly discuss the thin-wall limit, real-time corrections and extending the Wigner function framework to the high-temperature regime of EFTs.

If the temperature squared is much higher than the $N$ lowest eigenvalues, we can estimate the prefactor by expanding the $N$ lowest (hyperbolic) trigonometric functions in their arguments:
\begin{equation}
    \Gamma= \frac{V}{2\pi} \qty(\frac{\Delta E_\text{cb}}{2\pi T})^\frac{d}{2} \sqrt{\frac{\prod_{j=1}^N\lambda_{j}^\text{meta}}{\prod_{j=D+1}^N\lambda_j}}
    \qty(1+\sum_{j=1}^N \frac{\lambda_j^\text{meta}-\lambda_j}{24T^2} + \dots)
    \frac{\prod_{j>N}\sinh(\frac{\sqrt{\lambda_{j}^\text{meta}}}{2T})}{\prod_{j>N}\sinh(\frac{\sqrt{\lambda_{j_+}}}{2T})} e^{-\beta \Delta E_\text{cb}}\,.\label{eq:theResultHighT}
\end{equation}
Here, we have enumerated all the eigenvalues around the critical bubble as follows: $j=1$ is the negative eigenvalue, $j\in [2,\dots, d+1]$ are the translational zero values and $j\geq d+2$ are the positive eigenvalues. The approximated part of the prefactor is classical and matches with Langer's rate, correspondingly disagreeing with Linde's rate.

The correspondence of Eq.~\eqref{eq:theResultHighT} to the EFT approach~\cite{Gould:2021ccf} is quite straightforward to understand. The 4-dimensional Euclidean eigenvalues, $\lambda_{j\,2n}\sim \pi T$, defined in Eq.~\eqref{eq:EuclideanEigenvalues}, are integrated out into an effective free energy~\cite{Hirvonen:2022jba}:
\begin{equation}
    \Gamma= \frac{V}{2\pi} \qty(\frac{\Delta E_\text{cb}}{2\pi T})^\frac{d}{2} \sqrt{\frac{\prod_{j=1}^N\lambda_{j}^\text{meta}}{\prod_{j=D+1}^N\lambda_j}}
     \,e^{-\beta \Delta F_\text{eff,cb}}\,.\label{eq:highTenergySplit}
\end{equation}
All the fluctuations that are not constants in the Euclidean time dimension ($\abs{n}\geq2$ in Eq.~\eqref{eq:EuclideanEigenvalues}) are now integrated out. Hence, the remaining fluctuations around the critical bubble are described by a dimensionally reduced effective description~\cite{Farakos:1994kx,Kajantie:1995dw,Braaten:1995cm,Braaten:1995jr,Hirvonen:2022jba}, given by the effective free energy. (Note that the free energy is denoted by a three-dimensional action in the references.)

In the thin-wall limit, the classical behavior occurs earlier for the low-energy modes corresponding to wall deformations. These scale as $\lambda\sim R^{-2}$~\cite{Callan:1977pt} and thus are classical at large enough radii, $R\gg T^{-1}$, even at comparatively low temperatures, $T\sim m$. In 1+1 dimensions, there are no geometric wall deformations. Consequently, the only non-zero mode that can become classical at $T\sim m$ is the negative eigenvalue.

The two approaches of Wigner functions and high-temperature EFTs coincide in an interesting temperature regime: The temperature is high enough to generate the scale hierarchy, but low enough so that the backreaction from the $\lambda_{j\,2n}\sim \pi T$ modes onto the critical bubble and the $d$-dimensional eigenvalues can be treated perturbatively. At higher temperatures, the thermal effects become non-perturbatively strong, and the EFT construction becomes a necessity. We can show that this is the case for any-dimensional field theory -- noting that the temperature needs to be exponentially large in 1+1 dimensions.

The leading correction to the critical bubble at high temperatures from the nucleating scalar itself is given by the following diagram~\cite{Gould:2021ccf,Fukuda:1975di},\begin{equation}\label{eq:correctionToTheCriticalBubble}
    \feynalign{\criticalBubbleMod} \; \propto \; T^{D-2} \,,
\end{equation}
where the loop contains the high-energy modes of $\sim \pi T$ that modify the critical bubble, and the external leg corresponds to the low-energy modes of $\sim m$ that constitute the critical bubble. The temperature dependence follows from the high-energy modes. It only assumes the form of the kinetic term in the action in Eq.~\eqref{eq:aSimpleAction} and that the potential is just a perturbation in these temperatures for the high-energy modes. The cubic coupling in the diagram does not need to be present in the potential around the metastable phase as it will be generated by the nontrivial background, e.g.\ $\frac{\lambda}{4!}(\cb+\delta\phi)^4\supset\frac{\lambda\cb}{3!}\delta\phi^3$.

Even in 1+1 dimensions, the thermal correction to the critical bubble is enhanced by a logarithm, $\ln(T)$. Therefore, the leading perturbative correction to the critical bubble can always become non-perturbatively large at sufficiently high temperatures regardless of the dimension (\textit{cf.} the analysis below for concreteness).

The thermal correction to the quadratic term in the effective free energy follows the same scaling as the correction to the critical bubble:
\begin{equation}\label{eq:leadingMassCorrection}
    \feynalign{\fluctuationsMod}\;\propto\; T^{D-2}\,.
\end{equation}
This leads to the modification of the eigenvalues, $\lambda_j\to\lambda_j^\text{eff}$, which are given by the effective free energy.

To give some concreteness to the above EFT discussion, we can analyze the high-temperature corrections in the model given by an unbounded potential,
\begin{equation}
    V(\phi)=\frac{m^2}{2}\phi^2-\frac{\lambda}{4!}\phi^4\,,
\end{equation}
in 3+1 dimensions. The leading thermal correction to the effective potential is given by the diagram in Eq.~\eqref{eq:leadingMassCorrection}, leading to
\begin{equation}
    V_\text{eff}=\frac{m^2+\frac{\lambda T^2}{24}}{2}\phi^2-\frac{\lambda}{4!}\phi^4
\end{equation}
(see e.g.\ \cite{Laine:2016hma} or \cite{Hirvonen:2022jba}). The mass correction gives rise to the leading correction to the critical bubble. Hence, the nucleation rate formula in Eq.~\eqref{eq:theResult} does not require resummations and is valid without the EFT construction in temperatures of
\begin{equation}
    T^2\ll \frac{24 m^2}{\lambda}\,.
\end{equation}

In addition to the above, there are corrections that cannot be captured with equilibrium field theory. The corrections begin to appear on non-linear orders in the Wigner function approach. These include effects from off-equilibrium thermal particles, studied using Boltzmann equations in Ref.~\cite{Hirvonen:2024rfg}.

The new framework provides an avenue for constructing the real-time, high-temperature nucleation rate from first principles. This would include finding the above off-equilibrium particle effects, but also the effects from infrared quantum and fermionic fluctuations that have not yet been captured. For the full high-temperature results, a significant hurdle remains: generalizing the method to high temperatures. This could be possible for example with the framework of hard thermal loops~\cite{Pisarski:1988vd,Frenkel:1989br,Braaten:1989mz,Taylor:1990ia,Frenkel:1991ts,Braaten:1991gm}.

\section{Conclusions and outlook}\label{sec:conclusions}

In this article, we have presented a framework for computing the decay of metastable states in QFTs using Wigner functions. It is founded on the nonperturbative rate formula of the flux of the Wigner function, Eq.~\eqref{eq:nonperturbativeRate}, and the four steps \ref{i:step1}--\ref{i:step4} to evaluate it perturbatively.

In particular, we applied the framework to the over-the-barrier nucleation at intermediate temperatures, $T\sim m$. We first described the physical process on the CTP contour in Sec.~\ref{sec:quantumFieldNucleating}, and subsequently found the rate in Sec.~\ref{sec:equilibriumNucleation}, Eq.~\eqref{eq:theResult}. Notably, there are quantum effects in the prefactor that are not captured by Linde's rate, showing that the dynamics of the over-the-barrier nucleation can be quantum mechanical despite the classical critical bubble background. In addition, the formula asymptotes to Langer's rate through an EFT construction at high temperatures, rather than Linde's rate (Sec.~\ref{sec:highTemperatureLimit}).

Beyond purely theoretical interest, the simple rate result itself in Eq.~\eqref{eq:theResult} is already useful for analog experiments: The hierarchy between the temperature and mass is set by the experiment, and at least in Refs.~\cite{Zenesini:2023afv,Garcia:2025uph} it appears to coincide with $T\sim m$ rather than $T\gg m$, together with the classical temperature scaling of the exponent, $e^{-\Delta E_\text{cb}/T}$.

In cosmological phase transitions, the particular rate formula is unlikely to have observational consequences. This is due to the fact that it decreases exponentially in temperature, $\propto e^{-\Delta E_\text{cb}/T}$, and hence also in time. Since the nucleation rate starts from zero at the critical temperature, $\Gamma(T=T_\text{c})=0$, and is now decreasing, $\dd\Gamma(T\sim m) / \dd T<0$, it must have had a peak at higher temperatures, where thermal resummations are necessary. This peak likely produced exponentially more bubbles than the $T\sim m$ period -- thus making the $T\sim m$ period quite possibly unimportant. If the thermal period is not enough to complete the phase transition, it may complete via the constant vacuum decay rate at $T\ll m$.

The limit of validity of the rate result was bounded by the need for performing thermal resummations at high temperatures, which we showed to be the case for QFTs of any spacetime dimension in Sec.~\ref{sec:highTemperatureLimit}. The Wigner function analysis could potentially be extended to asymptotically high temperatures through the framework of hard thermal loops. There, it should reproduce the results from the Boltzmann equation analysis in Ref.~\cite{Hirvonen:2024rfg}, but also allow for capturing e.g.\ the effects of infrared quantum and fermionic fluctuations. It would also be interesting to see if the Wigner function approach gives a useful perspective on vacuum decay.

The perplexing mismatch between the singularity of the Wigner function at $T=\frac{\sqrt{\abs{\lambda_-}}}{\pi}$ and the rate formula at $T=\frac{\sqrt{\abs{\lambda_-}}}{2\pi}$, discussed in Sec.~\ref{sec:lowTValidity}, motivates theoretically going beyond the one-loop order. This would allow for distinguishing the actual low-temperature limit of validity -- whether the singular behavior of the Wigner function signals nonperturbative quantum effects onto the nucleation rate or is just illusory to the rate value. The two-loop analysis may also reveal interesting dynamical effects, as could coupling the nucleating field to gauge or fermionic fields.

\acknowledgments

We thank O.\ Gould for comments on the manuscript. The work was supported by the Royal Society Dorothy Hodgkin Fellowship, grant number DHF\textbackslash{}R1\textbackslash{}221001.

\appendix

\section{Aspects of the time evolution of the Wigner function}\label{app:WignerFunctionTimeEvolution}

Here, we will quickly obtain two aspects of the time evolution of the Wigner function:
\begin{enumerate}
    \item the non-perturbative flow for the Wigner function in Eq.~\eqref{eq:nonperturbativeflow},
    \item the approximate classical Liouville equation for the Wigner function on a quadratic potential for Eq.~\eqref{eq:CarvedOutWignerFunction}.
\end{enumerate}

Let us start from the von Neumann equation for the density matrix
\begin{equation}
    i\partial_t \rho[\phi,\phi']=\int\dd^d\mathbf{x}\qty(\frac{1}{2}\qty(-\funcdvsq{\phi}+\funcdvsq{\phi'}+(\grad\phi)^2-(\grad\phi')^2)+V(\phi)-V(\phi'))\rho[\phi,\phi']\,.
\end{equation}
In the retarded--advanced basis, Eq.~\eqref{eq:retardedAdvanceBasis}, this becomes
\begin{equation}
    i\partial_t \rho=\int\dd^d\mathbf{x}\qty(-\qty(\funcdv{\pr}\funcdv{\pa}+\pa\grad^2\pr)+V\qty(\pr+\frac{\pa}{2})-V\qty(\pr-\frac{\pa}{2}))\rho\,.
\end{equation}

We can integrate both sides with
\begin{equation}
    \int\mathcal{D}\pi\mathcal{D}\pa e^{-i \pi\cdot\pa}
\end{equation}
with the dot product defined to contain the spatial integral in Eq.~\eqref{eq:dotProduct}. The potential and gradient terms vanish due to the delta function $\delta(\pa)$ and the functional derivative $\funcdv{\pa}$ yields $i\pi$ under integration by parts. This leads to the non-perturbative time evolution of the Wigner function in Eq.~\eqref{eq:nonperturbativeflow} repeated here:\begin{equation}\label{eq:AppNonperturbativeflow}
    \int\mathcal{D}\pi\partial_t W[\pr,\pi] = - \int\mathcal{D}\pi\,\pi\cdot\funcdv{\pr} W[\pr,\pi]\,.
\end{equation}

For the Liouville equation, let us assume that the potential is quadratic and dependent on the spatial location,
\begin{equation}
    V(\phi)=\frac{m^2(\mathbf{x})}{2}\phi^2\,,
\end{equation}
as relevant to fluctuations around the critical bubble. The ``mass squared'', $m^2(\mathbf{x})$, can be negative. The von Neumann equation becomes
\begin{equation}
    i\partial_t \rho=\int\dd^d\mathbf{x}\qty(-\funcdv{\pr}\funcdv{\pa}+\pa\qty(-\grad^2+m^2(\mathbf{x}))\pr)\rho\,.
\end{equation}

We can integrate both sides with
\begin{equation}
    \int\mathcal{D}\pa e^{-i \pi\cdot\pa}
\end{equation}
We obtain the classical Liouville equation
\begin{equation}\label{eq:LiouvilleEquation}
    \partial_t W[\pr,\pi]=\int\dd^d\mathbf{x}\qty(-\pi\funcdv{\pr}+\Big(\qty(-\grad^2+m^2(\mathbf{x}))\pr\Big)\funcdv{\pi})W[\pr,\pi]\,.
\end{equation}

\section{One-loop equilibrium Wigner function on the critical bubble}\label{app:oneloopEquilibriumWignerFunction}

Here, we will explicitly perform the one-loop computation for finding the equilibrium density matrix and Wigner function around the critical bubble.

We begin from the Euclidean path-integral for the density matrix
(see e.g.\ Ref.~\cite{Laine:2016hma}):
\begin{align}
    \rho_\text{eq}[\phi,\phi']&=Z^{-1}\bra{\phi}e^{-\beta\hat{H}}\ket{\phi'}=Z^{-1}\int_{\pe(\tau=0)=\phi}^{\pe(\tau=\beta)=\phi'}\mathcal{D}\pe e^{-\se}\,,\\
    \se&=\int_0^\beta\dd\tau\int\dd^d\mathbf{x}\qty(\frac{1}{2}(\partial_\tau\pe)^2-\frac{1}{2}\pe\grad^2\pe+V(\pe))\,.\label{eq:fullEuclideanAction}
\end{align}

We can shift our integration variable, $\pe\to\cb+\pe$, to study the density matrix around the critical bubble, and expand the action to quadratic order:
\begin{align}
    \se&\approx\beta\overbrace{\int\dd^d\mathbf{x}\qty(-\frac{1}{2}\cb\grad^2\cb+V(\cb))}^{\equiv E_\text{cb}}\\
    &\qquad+\int_0^\beta\dd\tau\int\dd^d\mathbf{x}\frac{1}{2}\qty((\partial_\tau\pe)^2+\pe(-\grad^2+V''(\cb))\pe)\,.
\end{align}
The exponential suppression for the nucleation rate comes from the energy of the critical bubble if $\beta E_\text{cb}\gg1$.

We can continue the analysis by expanding the Euclidean field in the following orthonormal eigenbasis:
\begin{align}
    \pe=\sum_{j}a_{j}(\tau) f_j \,,\qquad (-\grad^2+V''(\cb))f_j=\lambda_jf_j\,.
\end{align}
This results in 
\begin{align}\label{eq:quadraticEuclideanAction}
    \se&\approx\beta E_\text{cb}
    +\sum_{j}\int_0^\beta\dd\tau\frac{1}{2}\qty((\partial_\tau a_j)^2+\lambda_j a_j^2)\,.
\end{align}

We can then expand the $a_j$s around the classical solution that satisfies the boundary conditions given by $\pr=\frac{\phi+\phi'}{2}$ and $\pa=\phi-\phi'$:
\begin{gather}
    a_j=a_j^\text{cl}+\delta a_j\,,\\
    \partial_\tau^2a_j^\text{cl}=\lambda_j a_j^\text{cl}\,,\qquad
    a_j^\text{cl}(0)=\int\dd^d\mathbf{x}f_j\phi\equiv\phi_j\,,\qquad
    a_j^\text{cl}(\beta)=\int\dd^d\mathbf{x}f_j\phi'\equiv\phi_j'\,,\label{eq:definingThecoefficients}\\
    \delta a_j(0)=\delta a_j(\beta)=0\,.
\end{gather}

The classical solutions are given by
\begin{align}
    a_j^\text{cl}&=\frac{\phi_j' \sinh(\sqrt{\lambda_j}\tau)+\phi_j \sinh(\sqrt{\lambda_j}(\beta-\tau))}{\sinh(\sqrt{\lambda_j}\beta)}\,,&&\lambda_j>0\,,\\
    a_j^\text{cl}&=\frac{\phi_j' \tau+\phi_j (\beta-\tau)}{\beta}\,,&&\lambda_j=0\,,\\
    a_j^\text{cl}&=\frac{\phi_j' \sin(\sqrt{-\lambda_j}\tau)+\phi_j \sin(\sqrt{-\lambda_j}(\beta-\tau))}{\sin(\sqrt{-\lambda_j}\beta)}\,,&&\lambda_j=\lambda_-\,.
\end{align}
Note that the latter two are analytic continuations of the first one. The analytic continuation only diverges unphysically for the negative eigenmode at $\lambda_-+\pi^2T^2=0$. This divergence is discussed further in Sec.~\ref{sec:lowTValidity}.

Integrating by parts, the Euclidean time derivatives to operate on the fluctuations in the quadratic Euclidean action in Eq.~\eqref{eq:quadraticEuclideanAction} and using the boundary conditions, we obtain
\begin{align}
    \se&\approx\beta E_\text{cb}
    +\sum_{j}\qty[\frac{1}{2}a_j^\text{cl}\partial_\tau a_j^\text{cl}\eval_{\tau=0}^{\tau=\beta}+\int_0^\beta\dd\tau\frac{1}{2}\delta a_j\qty(-\partial_\tau^2+\lambda_j)\delta a_j]\,.
\end{align}

Finally, we can expand $\delta a_j$ in an orthonormal basis that respects the boundary conditions:
\begin{equation}
    \delta a_j = \sum_{n\in \mathbb{Z}_+} \delta a_{jn} \sqrt{2T}\sin(n\pi T \tau)\,.\label{eq:eigenSpectrumOnTheInterval}
\end{equation}
The form of the quadratic Euclidean action becomes
\begin{align}
    \se&\approx\beta E_\text{cb}
    +\sum_{j}\qty[\frac{1}{2}a_j^\text{cl}\partial_\tau a_j^\text{cl}\eval_{\tau=0}^{\tau=\beta}+\sum_{n\in \mathbb{Z}_+}\frac{1}{2}\qty((n\pi T)^2+\lambda_j)\delta a_{jn}^2]\,.
\end{align}

We can now integrate over the fluctuations, which yields
\begin{align}\label{eq:equilibriumDensityMatrixOnCBinApp}
    \rho_\text{eq}[\cb+\phi,\cb+\phi']&=\frac{e^{-\beta E_\text{cb}}}{\tilde{Z}}\prod_j\sqrt{\frac{\beta \sqrt{\lambda_j}}{\sinh(\beta \sqrt{\lambda_j})}} \, e^{-\sqrt{\lambda_j}\tanh(\frac{\beta\sqrt{\lambda_j}}{2})\prj^2-\frac{1}{4}\sqrt{\lambda_j}\coth(\frac{\beta\sqrt{\lambda_j}}{2})\paj^2}\,.
\end{align}
Here, we have absorbed the eigenvalue-independent infinite products from the Jacobians and the Gaussian integrals into the normalization. We will confirm for the explicit one-loop Wigner function that it is normalized correctly. The variables $\prj,\paj$ are defined similarly to $\phi_j,\phi_j'$ in Eq.~\eqref{eq:definingThecoefficients}.

Now, we can perform the Wigner transformation of Eq.~\eqref{eq:WignerFunction} to get
\begin{equation}\label{eq:equilibriumWignerFunctionAppendix}
    W_\text{cb}[\cb+\pr,\pi]=\frac{e^{-\beta E_\text{cb}}}{\tilde{Z}}\prod_j\frac{1}{\sqrt{2}\cosh(\beta \sqrt{\lambda_j}/2)}\, e^{-\tanh(\frac{\beta\sqrt{\lambda_j}}{2})\qty(\sqrt{\lambda_j}\prj^2+\frac{\pi_j^2}{\sqrt{\lambda_j}})}
\end{equation}

The above result corresponds to the full equilibrium. However, for nucleation, we want our field to be in equilibrium in the metastable phase. This means that the normalization corresponds to the partition function around the metastable phase, $\tilde{Z}_\text{meta}$, instead of the total partition function, $\tilde{Z}$. 

Let us still find the normalization in the metastable phase to one-loop accuracy: We can find the partition function from the normalization condition of the Wigner function around the metastable state. Note that we have kept all the dependence on the potential through the eigenvalues and the energy. Hence, we can just read from above that
\begin{align}
    \tilde{Z}_\text{meta}&=e^{-\beta E_\text{meta}}\int\prod_j \dd \prj^\text{meta} \dd \pi_j \frac{1}{\sqrt{2}\cosh(\beta \sqrt{\lambda_j^\text{meta}}/2)} \, e^{-\tanh(\frac{\beta\sqrt{\lambda_j^\text{meta}}}{2})\qty(\sqrt{\lambda_j^\text{meta}}(\prj^\text{meta})^2+\frac{\pi_j^2}{\sqrt{\lambda_j^\text{meta}}})}\\
    &=e^{-\beta E_\text{meta}}\prod_j\frac{\pi}{\sqrt{2}\sinh(\frac{\beta\sqrt{\lambda_j^\text{meta}}}{2})} \label{eq:metastableNormalizationPartitionFunction}\,.
\end{align}
Note that we absorbed away one last Jacobian coming from changing from $\mathcal{D}\pr\mathcal{D}\pi$ to $\prod_j \dd\prj\dd\pi_j$. Hence, we must use the latter integration measure of the coefficients of the orthonormal basis functions when evaluating for the rate.

\section{On the truncated Wigner approach}\label{app:truncatedWigner}

Here, we will discuss the truncated Wigner approach used in Refs.~\cite{Calzetta:2001pp,Braden:2018tky,Hertzberg:2019wgx,Wang:2025ooq}. There, a classical equation of motion is constructed for the quantum field and the initial state is truncated to a Gaussian distribution relying on the smallness of the reduced Planck's constant, $\hbar$. Perturbative scattering calculations, for example, are often thought of as $\hbar$ expansions because each loop in the vacuum adds an additional power of $\hbar$. This works wonderfully well for quantum electrodynamics at low energies starting from the fundamental Lagrangian but not at all for quantum chromodynamics, despite $\hbar$ being the same for both theories. The answer to this conundrum is obvious: the perturbative expansion for scattering experiments is actually a coupling expansion, and the strong coupling is large at low energies. Similarly, one should be careful regarding the validity of the $\hbar$ expansion when studying the decay of metastable states.

First, we want to raise the issue already discussed in Ref.~\cite{Tranberg:2022noe}: the truncated Wigner approach (there called the classical-statistical approach) treats the system as a classical field theory that just happens to have non-equilibrium initial conditions. This includes the quantum zero-point fluctuations that are promoted into classical propagating and equilibrating fluctuations.

This has been discussed rather explicitly in Ref.~\cite{Wang:2025ooq}. First, they initialize the system with a Gaussian state with a Hartree-improved mass, $M_\text{H}$. For momentum fluctuations, this is given by
\begin{equation}
    \Big\langle\abs{\pi_\mathbf{k}}^2\Big\rangle=\omega_\mathbf{k}\qty(\frac{1}{2}+\frac{1}{e^{\beta \omega_\mathbf{k}}-1})\,,\quad \omega_\mathbf{k}=\sqrt{\mathbf{k}^2+M_\text{H}^2}\,.
\end{equation}
In order to compare to Langer's rate, they first need to construct an effective temperature based on the infrared momentum fluctuations,
\begin{equation}
    T_\text{eff}(t)=\frac{1}{N_\mathbf{k}}\sum_{\mathbf{k}<\Lambda} \abs{\pi_\mathbf{k}}^2,
\end{equation}
below a cut off $\Lambda$ given by the characteristic size of the bubbles. The construction is based on matching with the classical Rayleigh-Jeans distribution:
\begin{equation}
    \Big\langle\abs{\pi_\mathbf{k}}^2\Big\rangle\approx T_\text{eff}\,.
\end{equation}
The necessity of an effective temperature is fundamentally due to the classical equilibration taking place under the classical nonlinear equation of motion, and the evolution treating the zero-point fluctuations as out-of-equilibrium classical fluctuations. Based on Fig.~16, the quantity is also strongly time dependent. Our analysis however shows that the relevant temperature is the ``initial'' temperature, $T$, and not the effective temperature that changes in time due to an incorrectly truncated equation of motion.

Similarly, it seems that the system may need to equilibrate at zero temperature under the classical equation of motion for a few orders of magnitude more than the microscopic time scales before beginning to decay (see e.g.\ Fig.~5 in Ref.~\cite{Jenkins:2023eez}). One should really expect that the quantum nature of the initial Gaussian state is ruined by the classical time evolution, rather than promoted into a non-linear one properly containing quantum mechanical decay.

Let us then discuss the analytic approach in Ref.~\cite{Hertzberg:2019wgx}, which is more similar to our analysis. A classical equation of motion is derived from the real-time analysis similarly to Sec.~\ref{sec:quantumFieldNucleating}. The critical configuration would be the same critical bubble as ours in Eq.~\eqref{eq:criticalBubble}, but this is not really explicitly shown. On the basis of real-time considerations, a parametric estimate for the critical bubble is found in the thin-wall limit. The configuration of $d$ spatial dimensions is promptly promoted to a $d+1$ dimensional Euclidean time configuration with the same radius when evaluating the Euclidean action (\textit{cf.} Eq.~\eqref{eq:fullEuclideanAction} but with $\tau$ integration bounds from $-\infty$ to $\infty$). The Euclidean Coleman bounce is obtained from
\begin{equation}
    \qty(\partial_\tau^2+\grad^2)\phi_\text{b}=V'\qty(\phi_\text{b})\,.
\end{equation}
Hence, the radii of the $d$-dimensional critical bubble and the $d+1$-dimensional bounce are different. In general, the critical bubble is not a Euclidean-time slice of the bounce. 

The truncated Wigner approach can only really find nucleation that corresponds to the over-the-barrier process of the $d$-dimensional critical bubble, because it is based on classical equations of motion. In addition to the analytic discussion in Ref.~\cite{Hertzberg:2019wgx}, this is  exemplified in Fig.~2 of Ref.~\cite{Hertig:2026oav}. The top-left panel displays a great agreement between the critical configuration from the simulations and the tree-level critical bubble of the classical equations of motion, as expected. In the top-right panel, the cutoff is pushed higher, allowing for more modes to be populated by the vacuum zero-point fluctuations. Based on Ref.~\cite{Tranberg:2022noe}, we know that they correspond to just more classical fluctuations (potentially) out of equilibrium at the moment of nucleation. They should lead to radiative corrections onto the critical configuration (\textit{cf.} discussion around Eq.~\eqref{eq:correctionToTheCriticalBubble}), which is visible in the figure. The agreement is still very good (\textit{cf.} e.g.\ thermal radiative corrections in Fig.~6 in Ref.~\cite{Pirvu:2024nbe}).

The real-time equation is not really used to estimate the rate (apart from finding the critical radius), because the calculation from the Gaussian initial state is analytically intractable. Instead, the rate is estimated based on the configurations in the initial distribution that nucleate immediately. They obtain parametrically agreeing exponents from the initial-distribution and Callan-Coleman analysis. This may not be general, but holds in the examples given. Note that this still omits a dimensionless factor in the exponential, likely leading to exponentially disagreeing rates. From the numerical simulations, we can also see that the initial state does not generally accurately describe the decay rate from the real-time simulations. Also, the tail of the distribution, which in our analysis, is responsible for nucleation is truncated off from an initial Gaussian distribution.


 \newcommand{\noop}[1]{}

\providecommand{\href}[2]{#2}\begingroup\raggedright\endgroup

\end{document}